\documentclass{PoS}

\usepackage{amsmath} 

\newcommand{\bea}{\begin{eqnarray}}
\newcommand{\eea}{\end{eqnarray}}
\newcommand{\cN}{ \mathcal{N}} 
\newcommand{\cO}{\mathcal{O}} 
\newcommand{\mC}{\mathbb{C}} 
\newcommand{\tr}{ \hbox{tr}}

\title{  Permutations and the combinatorics of gauge invariants for general $N$
 }

\ShortTitle{ $N$-dependent Combinatorics of gauge theories. }

\author{\speaker{Sanjaye Ramgoolam }\\
       Centre for Research in string theory,\\ 
       School of Physics and Astronomy, \\ 
        Queen Mary, University of London \\
        E-mail: \email{s.ramgoolam@qmul.ac.uk}}

\abstract{ Group algebras of permutations have proved highly useful in solving a number of problems in 
large $N$ gauge theories.  I review the use of permutations in classifying gauge invariants 
in one-matrix and 
multi-matrix models and computing their correlators. These methods are also applicable to 
tensor models and have revealed a link between tensor models and the counting of branched covers. 
The key idea is to parametrize $U(N)$ gauge invariants using permutations, subject to equivalences. 
Correlators are related to group theoretic properties of these equivalence classes.  
Fourier transformation on symmetric groups by means of representation theory offers nice bases of functions on these equivalence classes. This has applications in AdS/CFT in identifying CFT duals of giant gravitons and their perturbations. It has also lead to general results on quiver gauge theory correlators, uncovering links to two dimensional topological field theory and the combinatorics of trace monoids. 
          }

\FullConference{Proceedings of the Corfu Summer Institute 2015 "School and Workshops on Elementary Particle Physics and Gravity"\\
		1-27 September 2015\\
		Corfu, Greece}

\begin{document}

\section{ Introduction  }

 The AdS/CFT correspondence \cite{Maldacena} relates ten dimensional 
 string theory to four dimensional conformal quantum field theory.
In CFT, the operator-state correspondence relates quantum states to 
local operators, which are gauge invariant composites built from the elementary fields. 
The enumeration of these composites and the computation of their correlators is 
an algebraic problem. The emergence from these algebraic structures of the extra dimensions of the string background, of the strings and branes in these backgrounds, which we expect from the duality is an example of geometry emerging from algebra.  The idea that space-time geometry  is to be reconstructed from algebras  underlies work in non-commutative geometry, a subject of interest to 
many in the audience at the Corfu2015 conference. In this talk, I will be describing some ways 
in which geometrical structures emerge from algebraic data in the context of CFT correlators.

The physical context of what I will describe is $ \cN =4 $ SYM, the canonical example 
of gauge string duality, where the dual is string theory in $ AdS_5 \times S^5$. 
The gauge group of the CFT4 is $  U(N)$. The fields transform in the adjoint. In addition to gauge fields 
and fermions,  we have six hermitian adjoint scalars $ X_1 , \cdots , X_6$. They transform 
in the vector representation  of an $ SO(6)$ global symmetry, which forms part of 
the $ PSU(2,2|4)$ superconformal algebra. 

Another algebra, more precisely a sequence of algebras, 
will play a crucial role in this talk. This is the sequence of group algebras of symmetric groups 
$ \mC ( S_n )$, for natural numbers $n $. Aside from the obvious Lagrangian symmetries,  
the direct sum  of symmetric group algebras, 
\bea 
\mC ( S_{ \infty} ) \equiv \bigoplus_{ n =0}^{ \infty } \mC ( S_n ) 
\eea
will be found to play an important role in the description of the state space of CFT. 
Permutation algebras are useful in organizing the multiplicities of representations of 
$ PSU(2,2|4)$. In the simplest sector of interest, the maximally supersymmetric half-BPS sector, 
the multiplicity of representations will be  related to problems of invariant theory of 
one matrix, and will be organized by the conjugacy classes of $S_n$. 
For the less supersymmetric quarter and eighth-BPS sectors,  we encounter a multi-matrix problem and 
a generalization of conjugacy class algebras will play a role.  

A very useful tool for thinking about the matrix invariant theory problems we encounter, 
and the role of permutations in these problems, will be tensor spaces and linear operators acting on 
 these tensor spaces. Diagrams for representing linear operators, their products and traces, provide a powerful way to uncover the hidden simplicity of multi-index tensor manipulations. These tools 
will be illustrated as we proceed from the simplest example of 1-matrix to more complex 
 multi-matrix problems 
encountered in AdS/CFT and to  problems of tensor invariants encountered in tensor models.

\section{ Half-BPS sector in $ \cN=4$ SYM: 1-complex matrix model } 

Using two of the six hermitian scalars in $ N =4$ SYM, define  $ Z = X_1 + i X_2 $. 
Gauge invariant operators such as 
\bea
\tr Z^3 ~~ , ~~  \tr Z^2 \tr Z ~~,~~  (\tr Z)^3
\eea
which are holomorphic gauge-invariant polynomials in the matrix elements $Z^i_{j}$
are annihilated by half of the 32 supercharges in the super-algebra $ PSU(2,2|4)$. 
Acting with the generators of the super-algebra produces an ultra-short representation. 
All ultra-short representations are generated in this way. Their correlators have non-renormalization properties (see \cite{BDP} and refs. therein to earlier work) which allows comparison between free conformal field theory computations and supergravity. These holomorphic gauge invariant operators 
satisfy the BPS condition $ \Delta = J $ relating the dimension to a $U(1) $ charge inside the $ SO(6)$ 
R-symmetry.  The enumeration of these BPS operators using a Young diagram basis, the computation of correlators in this basis and the identification of Young diagram states with giant gravitons  was developed in \cite{CJR}. 

\begin{figure}[ht]%
\begin{center}
\includegraphics[width=0.7\columnwidth]{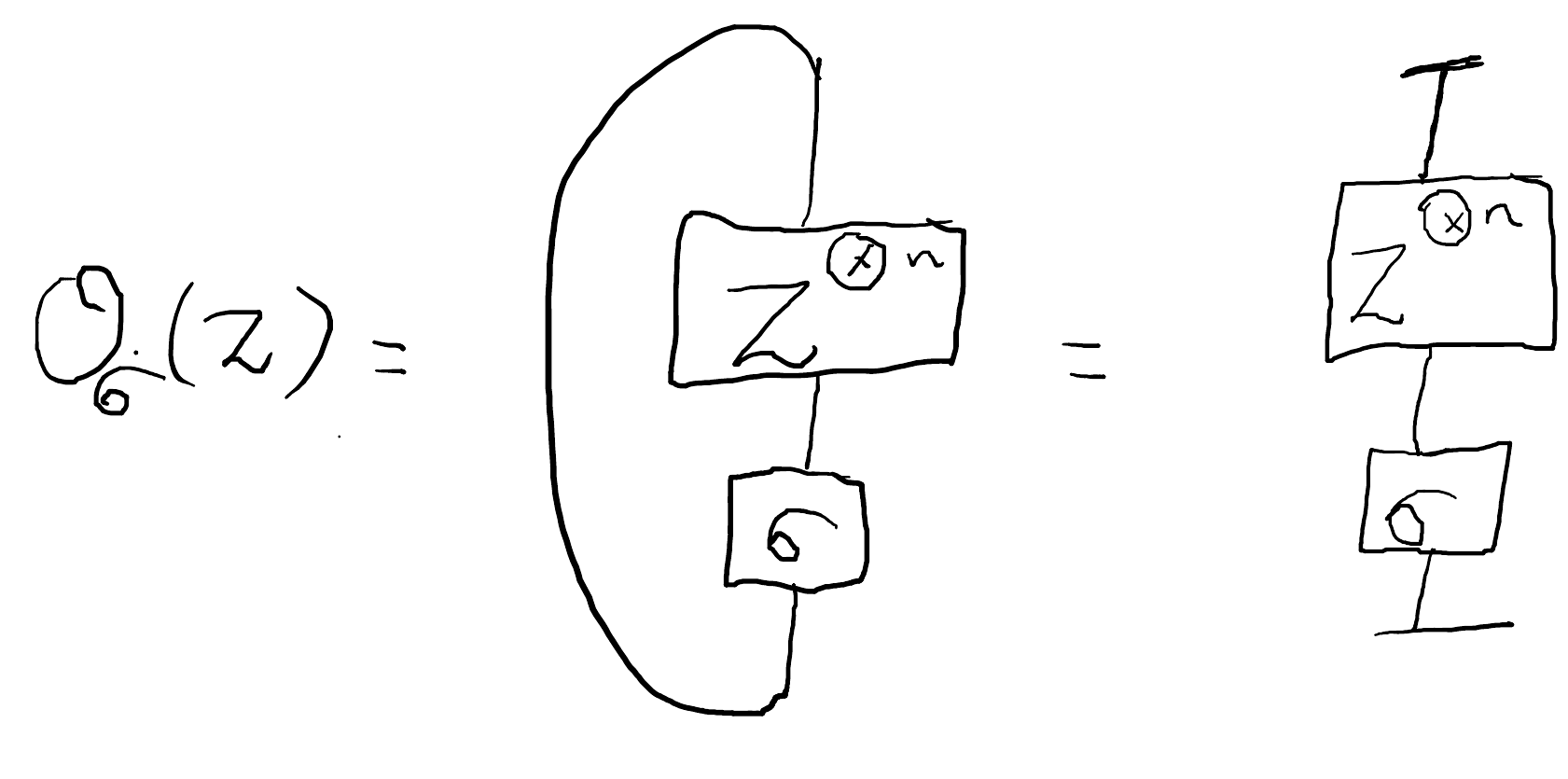}%
\caption{Operator as trace in tensor space   }%
\label{fig:OpD}%
\end{center}
\end{figure}

For a fixed $ \Delta = n $, and restricting to $ n < N $, the number of gauge invariant 
polynomials is equal to the number of partitions of $n$, i.e. the number of ways of writing 
$n$ as a sum of positive integers. This is also the number of conjugacy classes of the symmetric group 
and this is no accident. Gauge invariants can be constructed from permutations $ \sigma \in S_n$ 
which determine the order in which the lower indices are contracted with the upper indices
\bea\label{Osig} 
 \cO_{ \sigma } \equiv  Z^{i_1}_{i_{\sigma (1)}} \cdots Z^{i_n}_{ i_{\sigma(n)} } 
\eea
It is useful to think of $ Z $ as an operator acting on an $N$-dimensional vector space $V_N$
and $ Z^{ \otimes n } $ as an operator on the tensor product $V_N^{ \otimes n } $. 
There is a standard action of  $ \sigma  \in S_n $ on the $n$-fold tensor product. Choosing a set 
of basis vectors $ \{ e_i , 1\le i \le N \}$ for $V_N$, permutations act as 
\bea 
\sigma ~~ (  e_{ i_1 } \otimes e_{i_2} \cdots e_{i_n} ) = e_{ i_{ \sigma(1)} } \otimes \cdots  \otimes e_{ i_{ \sigma (n)} } 
\eea
The tensor product of $Z$-operators acts in the standard way as 
\bea 
Z^{ \otimes n }  ( e_{i_1} \otimes e_{i_2} \cdots \otimes e_{i_n} ) 
=  ( Z^{j_1}_{i_1} Z^{j_2}_{i_2} \cdots Z^{j_n}_{i_n} )  e_{j_1} \otimes e_{j_2} \cdots \otimes e_{j_n} 
\eea
One checks that 
\bea 
\cO_{ \sigma } ( Z )  && = \tr_{ V_N^{\otimes n } } ( Z^{\otimes n } \sigma ) \cr 
 && = \langle e^{ i_1} \otimes e^{ i_2} \otimes \cdots \otimes e^{ i_n } | Z^{ \otimes n } \sigma | e_{ i_1} \otimes e_{ i_2}  \cdots \otimes e_{ i_n} \rangle 
\eea
Using the standard connection between linear operators and diagrams, such as  used in knot theory for example, we can represent this diagrammatically as in Figure \ref{fig:OpD}.  In this picture a line 
represents a state in $V^{ \otimes n}_N$, a box is a linear operator and a contraction is an identification of an upper line with a lower line. On the left of the picture, this is represented  by joining top and bottom. On the right picture the upper index  line ends on a horizontal line, the lower index line 
ends on another horizontal line and the  two horizontal ines are understood to be identified. 
The diagrammatic way of thinking about operators and correlators is explained further in \cite{CR0205,QuivCalc}. 

It is  plausible and very easy to check that, for any permutation $ \gamma \in S_n$, 
\bea
\gamma Z^{\otimes n } \gamma^{-1} = Z^{ \otimes n } 
\eea
This is an equality of operators acting on $ V_N^{ \otimes n }$. 
It follows that 
\bea 
\cO_{ \gamma \sigma \gamma^{-1}  } ( Z ) = \tr_{ V_N^{ \otimes n } }  \left (  Z^{ \otimes n } \gamma \sigma \gamma^{-1} \right) = \tr_{ V_N^{ \otimes n } } \left ( \gamma^{-1} Z^{ \otimes n }\gamma \sigma \right ) = \cO_{ \sigma } ( Z ) 
\eea
The enumeration of gauge invariants is the same as enumerating conjugacy classes in $S_n$. 
Thus 
\bea 
\mC ( S_{ \infty} ) = \bigoplus_{ n } \mC ( S_n ) 
\eea
is a hidden algebra which organizes gauge invariants. 

\begin{figure}[ht]%
\begin{center}
\includegraphics[width=0.7\columnwidth]{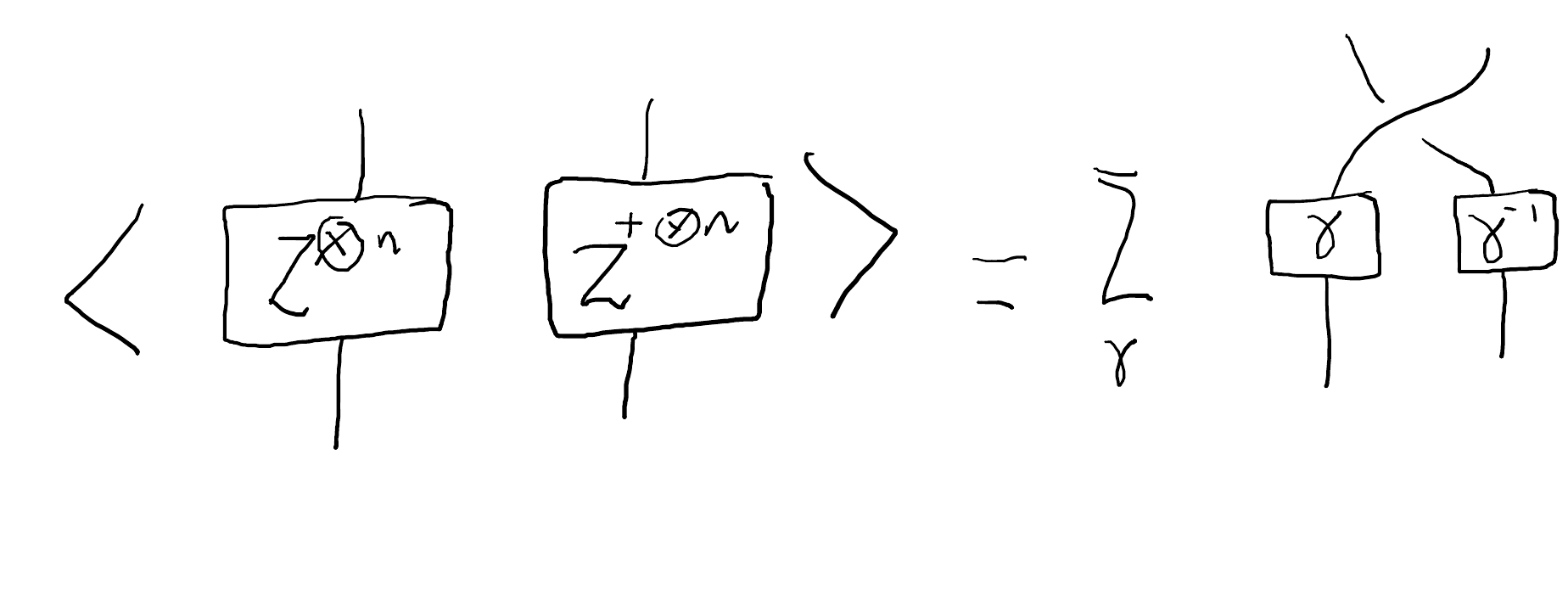}%
\caption{Wick contraction sum as operator in tensor space }%
\label{fig:Wick}%
\end{center}
\end{figure}

It turns out that permutations are important, not just for enumerating gauge invariants, but also 
for computing their correlators. Consider a two point function involving a holomorphic 
gauge invariant polynomial of fixed degree and an anti-holomorphic polynomial of fixed degree 
in the free field limit. This  is non-zero if the two polynomials have the same degree. The non-vanishing 
two-point functions are thus of  the form 
\bea 
\langle \cO_{ \sigma_{1} } ( Z ) ( x_1 )   ( \cO_{ \sigma_{2} } ( Z ) ( x_2) )^{ \dagger}  \rangle 
\eea
where $ \sigma_1 , \sigma_2 \in S_n$. The correlator is calculated using the basic correlator 
\bea 
 \langle Z^{i_1}_{j_1} ( x_1 ) (Z^{\dagger} )^{ i_2}_{j_2} ( x_2 ) \rangle =  { \delta^{i_1}_{j_2} \delta^{i_2}_{j_1} \over ( x_1 - x_2 )^2 } 
\eea
and Wick's theorem, which implies that 
\bea\label{wicksum-intens}
\langle Z^{i_1}_{j_1} \cdots  Z^{i_n }_{j_n} Z^{ \dagger k_1}_{ l_1} \cdots Z^{\dagger k_n }_{l_n} \rangle && = \delta_{j_{ \gamma (1 ) } }^{k_1}  \delta_{ l_1}^{ i_{ \gamma (1)} } \cdots
    \delta_{j_{ \gamma (n  ) } }^{k_n}  \delta_{ l_n}^{ i_{ \gamma (n )} } \cr 
&& = \delta_{j_{ \gamma (1 ) } }^{k_1}  \delta_{ l_{ \gamma^{-1} (1) }}^{ i_{ 1} } \cdots
    \delta_{j_{ \gamma (n  ) } }^{k_n}  \delta_{ l_{ \gamma^{-1} (n) }}^{ i_{ n } } 
\eea
We have dropped the space-time dependence which is trivial. 
Performing the free field path integral amounts to relacing the $ Z , Z^{\dagger} $ operators in tensor 
space by permutation operators. This can be expressed diagrammatically in Figure \ref{fig:Wick}.

 \begin{figure}[ht]%
\begin{center}
\includegraphics[width=0.7\columnwidth]{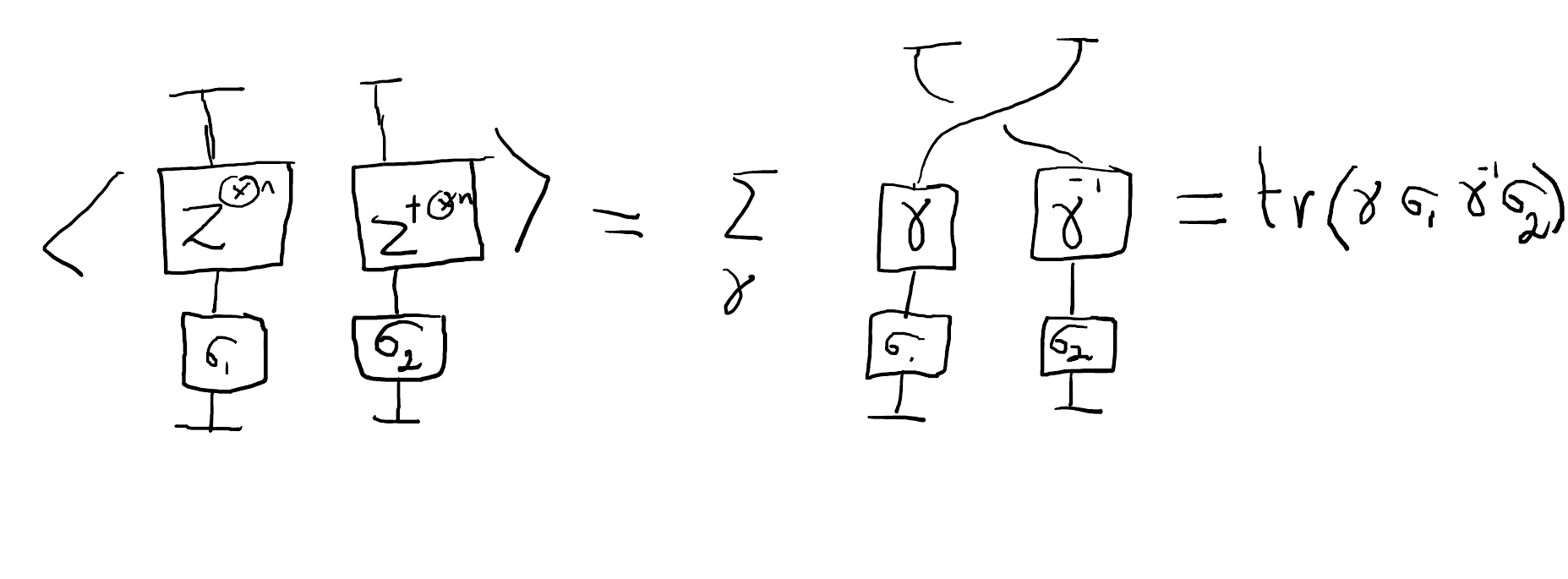}%
\caption{   Correlator as trace in tensor space  }%
\label{fig:CorTr}%
\end{center}
\end{figure}

The correlator can be computed by putting together (\ref{Osig}) and (\ref{wicksum-intens}). 
The calculation can also be understood diagrammatically as in Figure \ref{fig:CorTr}. 
The result is 
\bea 
{ ( x_1 - x_2)^{ 2n }}  \langle \cO_{\sigma_1  } ( Z ( x_1) ) ( \cO_{ \sigma_2} (Z (  x_2)  ) )^{\dagger}   \rangle 
&&  =  \sum_{ \gamma \in S_n } \tr_{ V_N^{ \otimes n } } ( \sigma_{1} \gamma \sigma_2^{-1}  \gamma^{-1} )   \cr 
&& = \sum_{ \gamma \in S_n }  N^{ C_{ \sigma_1\gamma \sigma_2^{-1}  \gamma^{-1} }} \cr 
&& = \sum_{ \gamma , \sigma_3 \in S_n  } N^{ C_{ \sigma_3}} \delta ( \sigma_1 \gamma \sigma_2 \gamma^{-1} \sigma_3 ) \cr 
&& = { n! \over |T_1| |T_2|}  \sum_{ \sigma_1' \in T_1, \sigma_2' \in T_2, \sigma_3 \in S_n } 
\delta ( \sigma_1' \sigma_2' \sigma_3 ) N^{ C_{ \sigma_3} }
\eea
The first step comes from the diagrammatic manipulation. In the second we use that, the trace in 
$V_N^{ \otimes n }$ of a permutation is given in terms of the number of cycles in $\sigma$, denoted by 
$ C_{ \sigma}$
\bea
\tr_{ V_N^{ \otimes n }}  ( \sigma ) = N^{ C_{\sigma}}
\eea
In the third line, we have introduced an extra permutation and a delta function on the permutation group to write a neater formula. $ \delta (\sigma )$ is defined to be $1$ if $ \sigma $ is the identity 
permutation and $0$ otherwise. Finally since $ \cO_{ \sigma}$ only depends on the conjugacy class, 
we have replaced $ \sigma_1, \sigma_2$ by sums of permutations in their respective 
conjugacy classes $ T_1 , T_2$, dividing out by the sizes of these conjugacy classes $ |T_1|, |T_2|$. 
With these sums in place, the $ \gamma$ sum can be replaced by $ n!$ since the sums commute with $ \gamma$. 

This shows that the introduction of permutations in $S_n$ 
as a way to parametrize the gauge invariant operators is extremely useful, 
since correlators can be given neatly in terms of group theory of $S_n$. 

The final formula has an interpretation in terms of counting of  branched covering maps from 
Riemann surfaces (worldsheets $ \Sigma$ ) to the sphere $ \mathbb{P}^1$  branched over three  points on the sphere. These branched covers are  holomorphic maps 
\bea 
f : \Sigma \rightarrow \mathbb {P}^1
\eea
with vanishing holomorphic derivatives at points on $ \Sigma $ which are in the inverse image of 
three points on the $\mathbb {P}^1$, which can be taken as $ \{ 0 , 1, \infty \} $
\cite{dMSR1002,Brown1009}.
These types of holomorphic maps are called Belyi maps and have deep implications for number theory, in connection with the absolute Galois group \cite{LandoZvonk}.

It is rather amazing that correlators in the distinguished half-BPS sector of $ \cN =4$ SYM, 
 equivalently in a distinguished $ AdS_5 \times S^5$ background of string theory,  are related directly to Belyi maps, of central  importance in number theory, a subject  which Gauss famously called  the Queen of Mathematics. 

The power of the permutation approach to gauge-invariants becomes particularly manifest 
when we consider finite $N$ effects. While the $ \cO_{ \sigma } ( Z ) $, with conjugation equivalence,  form a good basis of operators when $n \le N$ this is not the case for $n>N$. In this case, there are relations among traces due to the Cayley Hamilton theorem which implies  that $ tr Z^{ N+1}$ can be written in terms of multi-traces. An elegant way to give a basis at finite $N$ is to use the Fourier transform relation from group theory, which relates conjugacy classes to irreducible representations

Irreducible representations $R$ of $S_n$, correspond to Young diagrams with $n$ boxes. 
We can define operators corresponding to Young diagrams as 
\bea 
\cO_{ R } ( Z )   = { 1\over n! } \sum_{ \sigma \in S_n} \chi_R ( \sigma ) \cO_{ \sigma } ( Z ) 
\eea
where $ \chi_R ( \sigma )$ is the character of the $S_n$ group element $ \sigma $ in the 
irrep $R$, i.e. the trace of the matrix representing $ \sigma $ in irrep $R$. 
It is shown that  \cite{CJR}  these operators have diagonal 2-point function. 
\bea 
\langle \cO_{ R} ( Z ) (\cO_{ S } ( Z ) )^{\dagger} \rangle = \delta^{RS} { n! Dim R \over d_R } 
\eea
Failure of diagonality of the trace basis had been used \cite{BBNS0107} to argue that giant gravitons are not traces. Single giants which are large in the $S^5$ were identified with subdeterminant operators. 
With the Young diagram basis in hand, there are natural candidates for CFT duals of 
single and multi-giant graviton states \cite{CJR}.

\section{ Multi-matrix model }

The above approach to matrix invariants, based on permutations and diagrammatic tensor space methods, extends to multi-matrix models. This should be contrasted with many other approaches to 1-matrix models, e.g. reduction to eigenvalue dynamics,  which do not admit multi-matrix generalization. 
The enumeration and correlators  of multi-matrix invariants has direct applications to quarter and eighth BPS sectors of SYM. I will explain how permutation methods described above generalize to give 
restricted Schur operators. These  were introduced in the study of open string excitations of 
giant gravitons \cite{BBFH04,DSSI,DSSII,BDSIII}, subsequently shown to diagonalize the 2-point functions in the 2-matrix sector \cite{BCD0801} and understood in terms of enhanced symmetries 
in gauge theory \cite{KR3}.

Consider the two-matrix problem, of enumerating gauge invariants built from two matrices. 
In the context of the quarter-BPS sector of SYM, we are interested in holomorphic polynomial 
gauge invariants built from two matrices $ Z , Y$, where the gauge symmetry acts 
as 
\bea 
( Z , Y )  \rightarrow  ( UZ U^{\dagger} , U Y U^{ \dagger} ) 
\eea
As in the 1-matrix case, we observe that gauge invariants can be parametrized by permutations 
which determine how the lower indices are contracted with the upper indices 
\bea 
\cO_{ \sigma } ( Z , Y ) = \tr_{ V_{N}^{ \otimes m +n } }  \left ( Z^{ \otimes m }\otimes Y^{\otimes n } 
\sigma \right )  = Z^{ i_1}_{ i_{ \sigma (1)}} Z^{i_2}_{ i_{ \sigma(2)} } 
\cdots Z^{ i_{m}}_{ i_{ \sigma (m) } } Y^{ i_{m+1}}_{ i_{ \sigma (m+1)} } \cdots Y^{ i_{ m+n}}_{ i_{ \sigma (m+n)} }
\eea 
Here the permutation $ \sigma $ is in the symmetric group $S_{m+n}$, where $m$ is the 
number of $Z$ and $n$ is the number of $Y$ in the gauge-invariant operator. 
In this two-matrix case, the equivalence takes the form 
\bea\label{equivrest} 
\cO_{ \sigma  }  ( Z , Y ) = \cO_{ \gamma \sigma \gamma^{-1} }  ( Z , Y ) 
\eea
for $ \gamma \in S_m \times S_n$. 
In the large $N$ limit, i.e. when $ m+n < N$, these are the only equivalences. Counting 
gauge invariant operators is equivalent to enumerating orbits of $ S_m \times S_n$ permutations 
$ \gamma $ acting on permutations $ \sigma \in S_{ m+n}$ by conjugation : 
\bea
\sigma \sim  \gamma \sigma \gamma^{-1} ~~~~ \hbox{ for } ~~~~ \gamma \in S_m \times S_n 
\eea
From the Burnside Lemma,
we know that the number of orbits is equal to the number of fixed points, so that 
\begin{align*} 
\hbox{ Number of operators with $m,n$ copies of $Z , Y $ } = 
{ 1 \over m! n! } \sum_{ \gamma \in S_{ m } \times S_n } \sum_{ \sigma \in S_{m+n}} 
\delta ( \gamma \sigma \gamma^{-1} \sigma^{-1}  ) 
\end{align*} 
This formula can be used to obtain  the generating function 
\begin{align*} 
&& \sum_{ m , n = 0}^{ \infty} z^m y^n  
( \hbox{ Number of operators with $m,n$ copies of $Z , Y $ }  ) \cr 
&& = \prod_{ i =1}^{ \infty } { 1 \over ( 1 - z^i -  y^i ) } 
\end{align*} 
The expression also has an interpretation as a partition function 
for topological  lattice gauge theory with $ S_{m+n}$ gauge group on a two-torus, with a 
defect inserted on a circle, which constrains the holonomy to be in the $ S_m \times S_n$ subgroup. 
The free field 2-point function is 
\bea 
&& \langle \cO_{ \sigma_1  }  ( Z , Y )  ( \cO_{ \sigma_2  }  ( Z , Y )  )^{ \dagger} \rangle 
= \sum_{ \gamma } \tr_{ V_N^{ \otimes m+n } }  ( \sigma_1 \gamma \sigma_2 \gamma^{-1}  ) \cr 
&& = \sum_{ \gamma \in S_m \times S_n } \sum_{ \beta \in S_{m+n} } 
\delta  ( \sigma_1 \gamma \sigma_2^{-1}  \gamma^{-1} \beta ) N^{ C_{ \beta } } 
\eea

As in the 1-matrix case, finite $N$ counting and correlators can be obtained by 
going to a Fourier basis using matrix elements of the permutation $ \sigma \in S_{m+n}$ 
in irreps $R$. There are linear operators $D^R ( \sigma )$ which give functions
 $D^R_{ I J } ( \sigma )$ where $I,J$ can  be taken to run over an orthonormal basis for  the irrep $R$. Since the functions of interest are invariant under 
$S_m \times S_n$, it is useful to use group theoretic data associated with reduction of the 
irrep $R$ of $S_{m+n}$ into irreps of $ S_m \times S_n$. These can be parametrized by a
pair of Young diagram $ R_1, R_2 $ with $ m , n $ boxes respectively. A representation $R$ of 
$S_{m+n}$ can be decomposed into a direct sum of representations $R_1, R_2$, which appear 
with multiplicities $ g( R_1 , R_2 , R )$ equal to the Littlewood-Richardson coefficients. Thus there is a 
subgroup-adapted basis in $ V_R$, with states of the form 
\bea 
|R_1, R_2, m_1 , m_2 ; \nu_1 \rangle 
\eea
where $m_1, m_2$ are state labels for the irreps $R_1, R_2$ and $ \nu_1 $ is a multiplicity label 
taking values between $ 1 $ and $ g ( R_1  , R_2 , R )$. we may write the relevant decomposition as 
\bea 
V_{ R  }^{ ( S_{m+n} ) } = \bigoplus_{ R_1 , R_2 } V_{R_1}^{ ( S_m ) }  \otimes V_{R_2}^{ ( S_n ) } 
 \otimes V^R_{R_1, R_2} 
\eea
where the dimension of the multiplicity space $V^R_{R_1, R_2} $ is $ g ( R_1 , R_2 , R )$. 
The subgroup basis states can be expanded in terms of a general orthonormal basis 
\bea 
|R_1, R_2, m_1 , m_2 ; \nu_1 \rangle  = \sum_{ I } |  R , I  \rangle  \langle R , I |R_1, R_2, m_1 , m_2 ; \nu_1 \rangle 
\eea
The branching coefficients $ \langle R , I |R_1, R_2, m_1 , m_2 ; \nu_1 \rangle $ can be used 
to define functions $ \chi^R_{ R_1 , R_2 , \nu_1 , \nu_2 } ( \sigma )$, called restricted characters, 
labelled by three Young diagrams and two multiplicity indices. These functions are 
invariant under conjugation by $ \gamma \in S_{m } \times S_n$. These restricted characters 
are used to define restricted Schur Polynomials 
\bea\label{restschur}
\chi^R_{ R_1 , R_2 , \nu_1 , \nu_2 } ( Z , Y ) =
 \sum_{ \sigma \in S_{m+n}} \chi^R_{ R_1 , R_2 , \nu_1 , \nu_2 } ( \sigma )
\cO_{ \sigma } ( Z , Y ) 
\eea

In this Fourier basis, finite $N$ constraints are easy to implement and they follow from 
Schur-Weyl duality \cite{FulHar,SWrev}. They amount to restricting 
the  Young diagram $R$ to have no more than $N$ rows. Thus the finite $N$ counting of
2-matrix invariants is given by 
\bea && \hbox{ Finite $N$ counting of $2$-matrix gauge invariants } \cr 
&& \cr 
&& = \sum_{  \substack{   \hbox{ $R$ with  $m+n$  boxes} \\ { \hbox{$R$  has no more than  $N$ rows}}  }}   
\sum_{ \hbox{ $R_1$ with $m$  boxes }} \sum_{  \hbox{ $R_2$ with  $n$ boxes } } 
 ( g ( R_1 , R_2 , R )  )^2 \cr 
&& 
\eea
This counting formula can be obtained directly \cite{quivcalc} from the group integral formula
for counting gauge invariants which was introduced by Sundborg \cite{Sund9908}.
Another formula \cite{BHRI} for the counting in terms of $U(2)$ representations $ \Lambda $ is 
\bea && \hbox{ Finite $N$ counting of $2$-matrix gauge invariants } \cr 
&& \cr 
&& = \sum_{  \substack{   \hbox{ $R$ with  $m+n$  boxes} \\ { \hbox{$R$  has no more than  $N$ rows}}  }}   
\sum_{ \hbox{ $\Lambda $ with $m +n $  boxes }} 
 C ( R , R , \Lambda ) M ( \Lambda ; m , n )  \cr 
&& 
\eea
$ C ( R , R , \Lambda )$ is the $ S_n$  Kronecker coefficient (Clebsch-series multiplicity) for 
 $ \Lambda $ in $ R \otimes R$, and $ M ( \Lambda , m , n ) $ is the multiplicity of 
the irrep of the symmetric representations $ [m] \otimes [ n] $ of $ S_m \times S_n$ 
in the decomposition of $ \Lambda $ into irreps of $ S_m \times S_n$.  The equivalence 
of the two counting formulae was shown in \cite{collins}. 

The restricted Schur operators have diagonal 2-point function in the free field limit \cite{BCD0801}
\bea 
\langle \chi^R_{ R_1 , R_2 , \nu_1 , \nu_2 } ( Z , Y ) ( \chi_{S_1 , S_2 , \mu_1 , \mu_2 }^{ S }( Z , Y ) ) ^{\dagger} \rangle 
\propto \delta_{ R  , S } \delta_{ \nu_1 , \nu_2 } \delta_{ \mu_1 , \mu_2 } 
\eea

Diagonal bases in the 2-matrix problem, with well-defined $U(2)$ global symmetry quantum numbers
have also been constructed \cite{BHRI,BHRII}.  Brauer algebras provide another useful free field basis \cite{KR1}.
The operators in these different bases diagonalise commuting sets of Casimirs for 
enhanced symmetries in  the free field limit \cite{KR3}. For the two-matrix case, we have a 
 $ U(N) \times U(N)$ symmetry of left multiplication by unitary matrices 
\bea 
Z , Y \rightarrow U Z , VY 
\eea
We can also multiply on  the right. For each matrix there is in fact 
a $U(N^2)$ symmetry.
The Noether charges for these different symmetries can be used to construct Casimirs which measure 
the representation theory labels used to construct the different bases. Finding  minimal sets of charges determining the labels of the restricted Schur basis leads to the study of the structure of permutation centralizer algebras \cite{PCA}. The starting point of this discussion is that the 
averaging over the equivalence classes in (\ref{equivrest})
 and using the product in $ \mC ( S_{m+n} )$ 
to define a non-commutative associative algebra  $ \mathcal{A}  ( m , n ) $ on these equivalence classes. 
The restricted characters in \ref{restschur}  lead to 
 the Matrix (Wedderburn-Artin) decomposition of the algebra $ \mathcal{A} ( m ,n ) $ 
 in terms of  the expressions
\bea 
Q^{R}_{R_1, R_2 , \nu_1 ,\nu_2 } = \sum_{ \sigma \in S_{ m+n} } \chi^{R}_{R_1, R_2 , \nu_1 , \nu_2} ( \sigma ) ~~ \sigma 
\eea
This gives gives the intrinsic permutation meaning of the restricted characters, without reference to matrices. The structure of these algebras is closely related to Littlewood-Richardson coefficients, and can be 
viewed as a sort of categorification thereof. The relations between the centre and an appropriately defined Cartan (the diagonals of the Wedderburn Artin decomposition) play an important role in the 
definition of  minimal sets of charges.

\subsection{ Quivers} 

Recently these results have been generalized to the counting and correlators of 
quiver gauge theories, where the gauge group is a product of unitary groups
and  the matter is in bifundamentals or fundamentals \cite{quivcalc,quivwords,quivfuncor}.  Links to two dimensional  topological field theory have been described and some surprising connections to trace monoids which have applications in computer science have been described. This arises because the counting function 
for large ranks of  the Unitary groups is an infinite product of inverse determinants. 
The inverse determinant counts words constructed from an alphabet consisting of 
letters corresponding to simple loops, with partial commutation relations between the letters.

 \section{ Tensor models  } 
 
Consider a complex 3-index tensor $ \Phi_{ i , j , k } $
which transforms as $ V_{ N } \otimes V_{N} \otimes V_{ N } $
 and its conjugate $\bar \Phi^{ i , j , k  }$, which transforms as 
$ \bar V_N \otimes \bar V_N \otimes \bar V_N $. The three ranks can, more generally, be chosen 
to be unequal. Polynomial invariants are constructed by contracting  the upper indices with the lower indices \cite{GR1101}. 

  Permutation methods, of the type above which have been useful in enumerating multi-matrix invariants as outlined above, were used to approach the tensor invariants in \cite{BR1307}. Explicit generating functions for the counting, new connections with branched covers, as well as group theoretic expressions for correlators were found.

Choose a positive integer $n$ which determines the number of $ \Phi $ and $ \bar \Phi$ 
used to construct the invariants. These invariants can be parametrized by a 
triple of permutations $ ( \sigma_1 , \sigma_2 , \sigma_3 )$ which contract the $i$-indices, the $j$-indices and the $k$-indices. 
\bea 
\cO_{ \sigma_1 , \sigma_2 ,\sigma_3 } ( \Phi , \bar \Phi ) 
 = \bar \Phi^{ i_1 , j_1, k_1} \cdots \bar \Phi^{ i_n , j_n , k_n } 
     \Phi_{ i_{ \sigma_1(1)} , j_{ \sigma_2 (1)} , k_{ \sigma_3 (1) } } \cdots 
 \Phi_{ i_{ \sigma_1(n )} , j_{ \sigma_2 (n )} , k_{ \sigma_3 (n ) } }
\eea
There is an $S_n$ symmetry of permuting the $ \Phi $ and an $S_n$ symmetry 
of permuting the $ \bar \Phi$'s. Thus the counting of $3$-index tensor invariants of 
degree $n$ is equivalent to counting permutation triples  $ ( \sigma_1 , \sigma_2 , \sigma_3 )$ 
 subject to equivalences generated by $ \gamma_1 , \gamma_2 \in S_n$. 
\bea\label{equivtens}  
( \sigma_1 , \sigma_2 , \sigma_3 ) \sim ( \gamma_1 \sigma_1 \gamma_2 , \gamma_1 \sigma_2 \gamma_2 , \gamma_1  \sigma_3 \gamma_2 ) 
\eea
Using the Burnside Lemma, the number of these invariants is 
\bea 
{ 1 \over n!^2 } \sum_{ \gamma_1 , \gamma_2 \in S_n } \sum_{ \sigma_1, \sigma_2 , \sigma_3 } 
\delta ( \gamma_1 \sigma_1 \gamma_2 \sigma_1^{-1} ) \delta ( \gamma_1 \sigma_2 \gamma_2 \sigma_2^{-1} ) \delta ( \gamma_1 \sigma_3 \gamma_2 \sigma_3^{-1} ) 
\eea
This expression can be easily simplified. The first step is to solve for one of the $ \gamma$'s. 
The upshot is that we can write the counting as 
\bea 
{ 1 \over n! } \sum_{ \gamma \in S_n } \sum_{ \tau_1 , \tau_2 \in S_n } \delta ( \gamma \tau_1 \gamma^{-1} \tau_1^{-1} ) \delta ( \gamma \tau_2 \gamma^{-1} \tau_2^{-1} ) 
\eea
This is the counting of equivalence classes of pairs $ ( \tau_1 , \tau_2 ) $, where 
the equivalence is generated by simulataneous conjugation by a permutation $ \gamma$. 
This in turn is the same as counting branched covers of degree $n$ from Riemann surfaces to 
a sphere with three branch points. After choosing a base point on the sphere, labelling the inverse images as $ \{ 1 , \cdots , n \}$, and lifting paths going round the three branch points, we 
recover permutations $ ( \tau_1 , \tau_2 , (\tau_1 \tau_2)^{-1} ) $. The product of these three 
permutations is  the identity, as it should be, since the path surrounding all three branch points 
is homotopic to the tivial path, with trivial monodromy. Generalizing the above arguments shows 
that counting of tensor invariants of rank $d$ is equivalent to counting branched covers with $d$ 
branch points. This is used to construct explicit generating functions in \cite{BR1307}. 
As in the use of permutations for multi-matrix theories, the application 
to tensor models also shows that finite $N$ effects can be captured by 
performing the Fourier transform to representations. 

Averaging over the equivalence classes defined by (\ref{equivtens})  and taking the products 
in $ \mC ( S_n ) \otimes \mC ( S_n ) \otimes \mC ( S_n )$ leads to another important example of 
the permutation centralizer algebras defined in \cite{PCA} (see concluding section there).
The structure of this algebra is closely related to Kronecker product coefficients in $S_n$, 
forming a sort of categorification thereof.

\section{ Conclusions  and Outlook } 

I have described a body of techniques based on permutations and diagrams to approach 
problems of enumerating gauge invariants and computing correlators in
 1-matrix models, multi-matrix models, quiver gauge theories and tensor models.  
They are also useful in computing the spectrum of loop-corrected dilatation 
operators, particularly in the sector of perturbations around half-BPS 
giant gravitons  ( see \cite{DDJ1012,DMP1004,gigravosc,doubcos,BDT1602}  and references therein). 
Some progress has been made in finding the general quarter or eighth BPS ground states, 
 which are ground states of the 1-loop dilatation operator acting on holomorphic invariants
 built from two or three complex matrices \cite{PasRam1010,KimQuarter}. The general solution at finite $N$ remains elusive. One would like to construct the BPS operators in correspondence with 
states of a harmonic oscillator in two dimensions, as expected from quantization of 
giant graviton moduli spaces \cite{Mik0010,BGLM0606}. Generalizations beyond unitary groups have been studied \cite{CDD1301,CDD1303}. 

It is noteworthy that the algebra $ \mC ( S_{ n} ) $, in particular its centre, controls 
the $n$-dependent combinatorics of the orbifold theories. The $ \cN=(4,4)$ superconformal 
theory on the orbifold $ (T^4)^n/ S_{n } $ is dual to $AdS_3 \times S^3$ (some  papers 
 on the relevant combinatorics are \cite{Maldacena,deB9806,LM9906,JMR2000,LuMa0006,PRS0905}). 
This gives another instance where $ \mC( S_{\infty} ) $ knows about the interactions of gravitons in AdS   space, via holography. Elsewhere it has also been used in  the classification of brane-tilings \cite{JRR,HJRS,AHJPRR1104}, the counting of Feynman graphs \cite{FeynCount,RefCount}, 
the  systematic classification of light-cone string diagrams \cite{permLC}
and large $N$ 2d Yang Mills theory \cite{GT,CMR}. 

 The algebra $ \mC( S_{\infty} ) $, along with its representation theory 
and relations to Lie groups via Schur-Weyl duality, contains a lot of  information 
about holography and quantum field theory. Further examples and a deeper understanding of this phenomenon would be fascinating.

\begin{center} 
{\bf Acknowledgments} 
\end{center}

\vskip.3cm 

I thank the organizers of the Corfu2015 workshop on Non-commutative Field Theory and Gravity
for the opportunity to present this talk in September 2015 and the generous deadline extension
 allowed to complete the proceedings contribution.  I thank the audience for stimulating questions and feedback. I have taken the liberty to add a few small points in this article which draw on papers that have appeared since, but which fit well with the content of the talk. I am also grateful to all my collaborators on the research presented here: Joseph Ben Geloun, Tom Brown, Stefan Cordes, Steve Corley, David Garner, Laurent Freidel, Amihay Hanany, Yang Hui He, Paul Heslop, Robert de Mello Koch, Antal Jevicki, Vishnu Jejjala, Yusuke Kimura, Paolo Mattioli, Mihail Mihailescu, Greg Moore, Jurgis Pasukonis, Diego Rodriguez-Gomez, Rak Kyeong Seong. My research is supported by  STFC consolidated grant ST/L000415/1 ``String Theory, Gauge Theory \& Duality''

\end{document}